\documentclass[reprint,twocolumn,amsmath,amssymb,aps,superscriptaddress,prl]{revtex4-1}
\pdfoutput=1
\usepackage{graphicx}
\usepackage{bm}
\usepackage{epsfig}
\usepackage{amssymb}
\usepackage{amsfonts}
\usepackage{amsmath}
\usepackage{color}
\usepackage{epstopdf}
\usepackage{hyperref}
\usepackage{float}
\usepackage{appendix}
\restylefloat{table}
\usepackage{bibentry}
\usepackage{multirow}
\usepackage[caption=false]{subfig}
\usepackage{braket}
\newcommand{\ba}{\begin{eqnarray}}
\newcommand{\ea}{\end{eqnarray}}
\newcommand{\bd}{\begin{displaymath}}
\renewcommand{\v}[1]{{\bf #1}}
\newcommand{\nn}{\nonumber \\}
\DeclareGraphicsExtensions{.pdf,.png,.jpg}

\begin{document}
\title{Matrix Product Wave Function of the Ground State and Elementary Excitation \\ in the Spin-1/2 Chain}

\author{Jintae Kim}
\affiliation{Department of Physics, Sungkyunkwan University, Suwon 16419, Korea}
\author{Minsoo Kim}
\affiliation{Department of Physics, Sungkyunkwan University, Suwon 16419, Korea}
\author{Pramod Padmanabhan}
\affiliation{Department of Physics, Sungkyunkwan University, Suwon 16419, Korea}
\author{Jung Hoon Han}
\email[Electronic address:$~~$]{hanjemme@gmail.com}
\affiliation{Department of Physics, Sungkyunkwan University, Suwon 16419, Korea}
\author{Hyun-Yong Lee}
\email[Electronic address:$~~$]{hyunyong@korea.ac.kr}
\affiliation{Department of Applied Physics, Graduate School, Korea University, Sejong 30019, Korea}
\affiliation{Division of Display and Semiconductor Physics, Korea University, Sejong 30019, Korea}
\affiliation{Interdisciplinary Program in E$\cdot$ICT-Culture-Sports Convergence, Korea University, Sejong 30019, Korea}

\begin{abstract} We present variational matrix product state (vMPS) for the ground state of the spin-1/2 Heisenberg model. The MPS effectively organizes the various dimer configurations, in faithful reflection of the resonating valence bond (RVB) picture of the spin liquid, with the energy only 0.024\% higher than the exact value given by Bethe ansatz. Building on the ground-state vMPS, the one-spin wave function is constructed in a simple manner with the dispersion that matches well with the exact spectrum. The vMPS scheme is applied to the family of Hamiltonian extrapolating between the Heisenberg model and the Majumdar-Ghosh model. 
\end{abstract}
\maketitle
%

The idea of resonating valence bond (RVB) as a description of the spin liquid phase in low dimensions was forcefully advanced by Anderson~\cite{anderson87}, and has since played a central role in the study of this exotic phase of spin matter. The explicit construction of variational RVB state in Ref. \cite{anderson88} has added credibility to Anderson's original idea. Although not often stated as such, the ground state of the one-dimensional antiferromagnetic Heisenberg spin-$1/2$ chain, first solved by Bethe's ansatz, is a famous example of spin liquid phase in low-dimensional spin systems. Though the Bethe ansatz solution is by itself absolutely accurate, the ground state wave function obtained in that way (essentially as a collection of plane waves) fails to capture the spin liquid picture immediately. Attempts were made in the past to derive the ground state of the Heisenberg model in the RVB picture~\cite{oguchi89,saito90}, though the idea did not seem to have gained much traction since then. With the advance of tensor network method in recent years, increasing efforts to write down spin liquid wave functions in the tensor network form have been made~\cite{Didier12,Schuch12,Shenghan15,Dong18,Didier19,Schuch19,Schuch20}. We propose the one-dimensional tensor network - known as the matrix product state (MPS) - version of the variational ground state for the Heisenberg spin-1/2 chain that builds faithfully on the RVB picture of the superposition of dimer configurations and with excellent energy and other physical properties. The validity of the variation MPS (vMPS) construction is further supported by the construction of the spinon wave function in the same framework, and a good match of its energy dispersion against the exact result. The spin liquid picture becomes transparent in the vMPS construction. 

It was proven that the entire collection of non-crossing dimers spans the Hilbert space of total spin singlets for spin-$1/2$ chain~\cite{oguchi89,saito90}, to which the ground state must belong. A dimer $[ij]$ refers to a spin singlet $(|\uparrow_i \downarrow_j \rangle - |\downarrow_i \uparrow_j \rangle )/\sqrt{2}$ over a pair of $(i,j)$ sites in the lattice. A ``parent" dimer configuration is the $[12]\cdots[L-1, L]$ (or $[23][34]\cdots [L, 1]$ for a closed chain) for even lattice sites $L$. Applying permutations on the site indices $(1, \cdots, L)$ gives rise to new dimer configuration $[P1, P2][P3, P4]\cdots$. Some of these permutations generate crossed dimers, such as $[13][24]$, which breaks down to two non-crossing dimers by virtue of the easily derived identity: $[13][24] = [12][34] + [14][23]$. The identity can be extended to say that an arbitrary crossed dimer state can be decomposed as a sum of several non-crossing dimer configurations. Furthermore, one does not have the possibility of trimers or any singlets out of an odd number of spin-$1/2$'s. The tetramer [1234] breaks down as the sum  $[1234] = [12][34]+2[14][23]$, and the same fate awaits all bigger $2q$-mers. In short, vMPS only needs to capture the various dimer configurations in some efficient manner. 

Our strategy in constructing a good vMPS for the spin-1/2 chain is to start with the well-known Majumdar-Ghosh (MG) model and its exact ground state~\cite{MG}
\ba H_{\rm MG} = \sum_i \Bigl( {\v S}_i\cdot {\v S}_{i+1} + \frac{1}{2} {\v S}_i\cdot {\v S}_{i+2} \Bigl) . \ea
Each $\v S_i$ is a spin-$1/2$ operator at the lattice site $i$. The ground state spontaneously breaks the translation symmetry by forming $[2n, 2n+1]$ dimers which are compact dimers for all integers $1\le n \le L/2 -1$, or $[2n-1, 2n]$ dimers. The even and odd MG states, $|{\rm MG}_e \rangle$ and $|{\rm MG}_o \rangle$, can be linearly combined to restore translational symmetry: $ |{\rm MG} \rangle = |{\rm MG}_{e} \rangle + |{\rm MG}_o \rangle$. The symmetrized MG state $|{\rm MG} \rangle$ has a MPS representation in terms of the site ($T^s$) and bond $(B)$ tensors: $\left[T_1 \right]_{ij}^s = \delta_{i0} \delta_{js}$, $\left[T_1' \right]_{ij}^s = \delta_{is} \delta_{j0} , B_{ij} =  ( 1 \oplus [{\rm CG}_{\frac{1}{2}\frac{1}{2}}^0 ] )_{ij}$. 
The site tensors depend on the physical ($s =1,2$) and the virtual ($0 \le i,j \le 2$) indices. The bond tensor depends only on the virtual indices, which span the spin-$0$  ($i, j = 0$) and the spin-$1/2$ sector ($i,j=1,2$), for a total bond dimension $D=3$. The Clebsch-Gordon (CG) notation $[{\rm CG}^s_{s_1 s_2} ]$ for fusing two spins $(s_1, s_2)$ into the spin-$s$ is used throughout the paper. One can find its explicit form in the Supplementary Materials (SM).

\begin{figure}[h]
\includegraphics[width=0.45\textwidth]{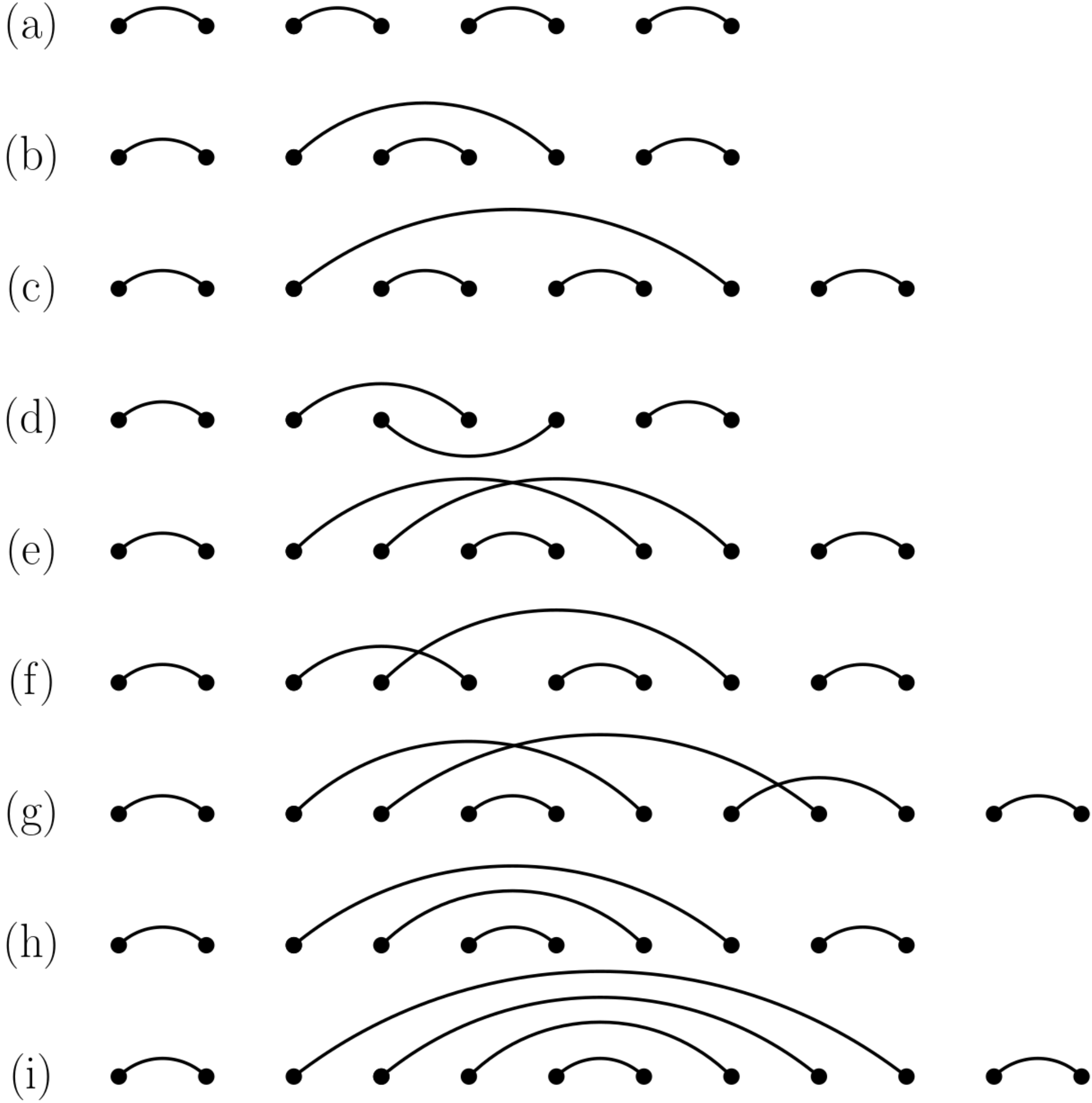} 
\caption{ (a)-(d) dimer configurations of $\{T_1,T_1',T_2,T_2'\}$ (c) an extended dimer arching over the two compact dimers (d) crossed dimers (e)-(f) complex dimer configurations of the full tensor combination (h) a 3-rainbow dimer configuration (i) a 4-rainbow dimer configuration.} \label{fig:1}
\end{figure}

The MPS wave function is obtained by first making the matrix product of the site tensor with the bond tensor $T^s B \equiv \overline{T}^s$: $\left[ \overline{T}_1 \right]_{ij}^s = \delta_{i0} [{\rm CG}_{{1\over2}{1\over2}}^0 ]_{sj},~\left[\overline{T}_1' \right]_{ij}^s = \delta_{is} \delta_{j0}$. The index $j$ runs over $j=1,2$ in $[{\rm CG}_{{1\over2}{1\over2}}^0 ]_{sj}$. With the site tensor given by $T = T_1 + T_1'$, one can show  
\ba \sum_{ \{ s \} } {\rm Tr} [ \overline{T}^{s_1} \overline{T}^{s_2} \cdots \overline{T}^{s_N} ] | \{ s \} \rangle = |{\rm MG}_{e} \rangle + |{\rm MG}_o \rangle , \label{eq:2}  \ea
$\{s\} = ( s_1, \cdots, s_N )$. One can show  $\overline{T}_1^{s_i} \overline{T}_1^{s_j} = 0 = \overline{T_1'}^{s_i} \overline{T_1'}^{s_j}$, while $[ \overline{T}_1  \overline{T}_1' ]^{s_1 s_2}_{ij} = \delta_{i0} \delta_{j0} [{\rm CG}_{{1\over2}{1\over2}}^0 ]_{s_1 s_2 }$ produces a compact dimer. The MG state has the energy $\langle \v S_i \cdot \v S_{i+1} \rangle = -3/8 = -0.375$,  far higher than the exact ground state energy of the Heisenberg model from the Bethe ansatz, $E_{\rm BA} =1/4 - \ln 2 = -0.443147$. The MG state represents an ordered array of dimers with zero inter-dimer correlations as shown in Fig. \ref{fig:1}(a). We can  {\it improve} upon the MG state by incorporating more and more of the resonant dimer states shown in Fig. \ref{fig:1}.

Generalizing the MG tensor, we propose the following $D=36$ site tensor $T=T_1+T_1'+T_2+T_2'+T_3+T_3'+T_4+T_4'$ with 23 parameters. Actual MPS is obtained by first multiplying $T^s$ and $B$ to get $\overline{T}^s = T^s B$, then taking their product as in Eq. (\ref{eq:2}). Explicit forms of $T_i$'s are
\ba
\left[T_1 \right]^s_{ii', jj'}  &=&  \delta_{i0} \delta_{js} [ \chi_1 ]_{i'j' } \quad~~~
\left[T_1' \right]^s_{ii', jj'}  =  \delta_{is} \delta_{j0} [\chi_{1'}]_{i' j'}\nn
\left[ T_2 \right]^s_{ii', jj' } &=& [\chi_2]_{ij} \delta_{i' 0 } \delta_{j' s} \quad~~~ \left[ T_2' \right]^s_{ii', jj' } = [\chi_{2'}]_{i j}\delta_{i' s } \delta_{j' 0} \nn
\left[ T_3 \right]^s_{ii', jj' } &=&  [ {\rm CG}_{ {1\over 2} 1}^{ {1\over 2}}  ]^s_{i j} [\chi_3]_{i' j'}~~ \left[ T_3' \right]^s_{ii', jj' } =  [ {\rm CG}_{1 {1\over 2}}^{ {1\over 2}}  ]^s_{i j} [\chi_{3'}]_{i' j'}\nn
\left[ T_4 \right]^s_{ii', jj' } &=& [\chi_4]_{i j} [ {\rm CG}_{{1\over 2} 1}^{{1\over2}} ]^s_{i'j' }~~ \left[ T_4' \right]^s_{ii', jj' } = [\chi_{4'}]_{i j} [ {\rm CG}_{1 {1\over 2} }^{{1\over2}} ]^s_{i'j' }\nn
\left[\chi_{k}\right]_{ij} &=& \left( x_{k,0} \oplus x_{k,\frac{1}{2}} [ {\rm CG}_{ {1\over 2} {1\over 2}}^0] \oplus x_{k,1} [ {\rm CG}_{ {1} {1}}^0] \right)_{ij} . \label{eq:fullT}
\ea
The virtual indices span the spin-$0$  ($i, j, i', j' = 0$), the spin-$1/2$ ($i,j, i', j'=1,2$), and the spin-$1$ sector ($i,j, i', j'=3,4,5$). The CG tensor is defined within their appropriate subspaces. Each $\chi_k$-tensor is parametrized by three variables $x_{k,0}, x_{k,{1\over2}}, x_{k,1}$ denoting the weights of the virtual spin-0, spin-1/2, and spin-1 sector, respectively. The bond tensor is symmetric in the two sets of virtual indices: $B_{ii', jj' } =  ( 1 \oplus  [ {\rm CG}_{ {1\over 2} {1\over 2}}^0] \oplus  [ {\rm CG}_{1 1}^0 ])_{ij} ( 1 \oplus  [ {\rm CG}_{ {1\over 2} {1\over 2}}^0] \oplus  [ {\rm CG}_{1 1}^0 ] )_{i'j'}$.
The first weight $x_{1,0}$ can be set to 1 without loss of generality. Optimization over the remaining 23-parameter space gives the energy $E=\langle \v S_i \cdot \v S_{i+1} \rangle = -0.44304007$, only about $0.024\%$ higher than the exact energy. The vMPS calculations were performed on $L=2000$ lattice using periodic boundary conditions. There is little discernible change in the variational energy for $L\gtrsim 1000$.

These tensors possess certain symmetries. Interchanging the two sets of virtual indices $(ij) \leftrightarrow (i'j')$ brings $T_1, T_1', T_3, T_3'$ to  $T_2, T_2', T_4, T_4'$, respectively, implying that identical structures can be generated from employing $\{ T_2, T_2', T_4, T_4' \}$ tensors as from $\{ T_1, T_1', T_3, T_3' \}$. By employing all eight tensors, though, one can generate a richer variety of dimer configurations. First of all, $\{ T_1 ,  T_1' \}$ alone reproduce the MG state [Fig. \ref{fig:1}(a)]. The $\{ T_1, T_1', T_2, T_2' \}$ tensors allow a certain subset of configurations such as (i) compact dimers [Fig. \ref{fig:1}(b)], (ii) extended dimers $[m,n]$ with $m+1 < n$ which encloses other compact dimers [Fig. \ref{fig:1}(c)], and (iii) crossed dimers $[ik][jl]~(i<j<k<l)$ [Fig. \ref{fig:1}(d)]. Some dimer structures can be expressed in multiple ways. Both $\overline{T}_1 \overline{T}_2 \overline{T}_2' \overline{T}_1'$ and $\overline{T}_2 \overline{T}_1 \overline{T}_1' \overline{T}_2'$ give $[14][23]$, for instance. The $\{ T_1, T_1', T_3 , T_3' \}$ tensors generate all the structures already given by $\{ T_1, T_1', T_2 , T_2' \}$, except the crossed dimers. 
The crossed dimer configuration [13][24] is equal to the sum of two non-crossing dimer [12][34] + [14][23]. The argument can be applied to higher-order crossed dimers such as [14][25][36]. In the end, all crossed dimer configurations decompose into various non-crossing dimers. The full tensor combination $\{ T_1, T_1', T_2, T_2', T_3, T_3',T_4,T_4' \}$ can give rise to additional structures such as (i) an extended dimer enclosing crossed dimers (e.g. [16][24][35]), (ii) more complex, crossed structure such as $[15][26][34]$, $[13][26][45]$, $[15][27][34][68]$ [Fig. \ref{fig:1}(e),(f),(g)], (iii) more complex ``rainbow" structures like $[16][25][34]$ and $[18][27][36][45]$ (Fig. \ref{fig:1}(h),(i)]. For example, $[15][27][34][68]$ and $[18][27][36][45]$ come from the products $\left[T_1 T_2 T_4 T_4' T_3 T_3' T_2' T_1'\right]$ and $\left[T_1 T_2 T_3 T_4 T_4' T_3' T_2' T_1'\right]$, respectively.

The $D=36$ tensor can capture the $n$-rainbow structure given by $[1,2n][2,2n-1]\cdots[n-1,n]$ for $n \le 4$ but not for $n>4$. By examining the line of thinking that led us from the $D=3$ MG tensor to the $D=36$ tensor, we can also figure out a way to go to bigger tensors. One way to enlarge the tensor space is to go from e.g. $0\oplus \frac{1}{2}$ to $0\oplus \frac{1}{2} \oplus 1$ in the virtual spin space. Such extension gives rise to $D=6$ tensor $T = T_1 + T_1' + T_3 + T_3'$, 
\ba
[ T_1 ]^s_{ij} & = &   \delta_{i0} \delta_{js}~~~~~\left[ T'_1 \right]^s_{i, j} =   \delta_{is}  \delta_{j0} \nn
\left[ T_3 \right]^s_{ij } & = &  [ {\rm CG}_{ {1\over 2} 1}^{ {1\over 2}}  ]^s_{i j}~~\left[ T'_3 \right]^s_{ij}  =  a [ {\rm CG}_{ 1 {1\over 2}}^{{1\over2}} ]^s_{ij } \nn 
B_{ij}  & = &   ( 1 \oplus  [ {\rm CG}_{ {1\over 2} {1\over 2}}^0 ] \oplus  [ {\rm CG}_{1 1}^0 ] )_{ij} , 
\label{eq:D6}  \ea
with one adjustable parameter $a$.
%
%
%
One can show 
%
%
\ba
[\overline{T}_1&& (\overline{T}_3 \overline{T}_3')^{q-1} \overline{T}_1']\sim [{\rm CG}_{ {1\over 2} {1\over 2}}^{1} ]_{s_1 s_2}^{\alpha} [{\rm CG}_{ 1 {1\over 2}}^{{1\over2}} ]_{\alpha s_3}^{\beta} [{\rm CG}_{ {1\over 2} {1\over 2}}^{{1\over2}} ]_{\beta s_4}^{\gamma}\nn
&&[{\rm CG}_{ 1 {1\over 2}}^{{1\over2}} ]_{\gamma s_5}^{\delta} [{\rm CG}_{ {1\over 2} {1\over 2}}^{{1\over2}} ]_{\delta s_6}^{\epsilon} \cdots [{\rm CG}_{ 1 {1\over 2}}^{{1\over2}} ]_{\zeta s_{2q-1}}^{\eta} [{\rm CG}_{ {1\over 2} {1\over 2}}^{0} ]_{\eta s_{2q}}
\ea
represents a $2q$-mer, i.e. a singlet made from $2q$ adjacent spin-1/2's. As  mentioned before, all $2q$-mers break down as the product of $q$ dimers, e.g. 
%
\ba
\left[\overline{T}_1 \overline{T}_3 \overline{T}_3' \overline{T}_1'\right] & \sim & [12][34]+ 2 [13][24]\nn
\left[\overline{T}_1 (\overline{T}_3 \overline{T}_3')^2 \overline{T}_1'\right] & \sim & [12][34][56]+ 2[12][36][45] \nn
&& + 2[14][23][56]+ 4 [16][23][45]. 
\ea
The introduction of the spin-$1$ virtual sector is a convenient device for writing down $2q$-mers efficiently, which in turn is an efficient way of generating diverse dimer configurations. The second approach is to make multiple copies of the virtual indices, e.g. two copies of $(1\oplus \frac{1}{2})$ virtual spaces for $(ij)$ and $(i'j')$ bond indices. This way of extending the tensor can give rise to $D=9$ tensor $T=T_1+T_1'+T_2+T_2'$ with 
\ba [T_1 ]^s_{ii', jj'}  &=&  \delta_{i0} \delta_{js} [ \chi_1 ]_{i'j' }~~~~ \left[ T_1' \right]^s_{ii', jj' }  =  \delta_{is} \delta_{j0}  [ \chi_{1'} ]_{i'j' }  \nn
\left[ T_2 \right]^s_{ii', jj' } &=& [ \chi_2 ]_{ij } \delta_{i' 0 } \delta_{j' s}~~~~ \left[ T_2^t \right]^s_{ii', jj' } = [ \chi_{2'} ]_{ij }  \delta_{i' s} \delta_{j' 0} \nn
\left[\chi_{k}\right]_{ij} &=& ( x_{k,0} \oplus x_{k,\frac{1}{2}} [ {\rm CG}_{ {1\over 2} {1\over 2}}^0] )_{ij} \nn
B_{ii', jj' } &=& ( 1 \oplus  [ {\rm CG}_{ {1\over 2} {1\over 2}}^0 ] )_{ij} ( 1 \oplus  [ {\rm CG}_{ {1\over 2} {1\over 2}}^0 ] )_{i'j'} .
\label{eq:D9} \ea
%
The $D=36$ tensor given in Eq. (\ref{eq:fullT}) incorporates both ways of extending the tensor. For completeness, we write down the $D=100$ tensor construction from the $0\oplus \frac{1}{2} \oplus 1 \oplus \frac{3}{2}$ virtual space in SM. 

\begin{figure}[htb]
\includegraphics[width=0.48\textwidth]{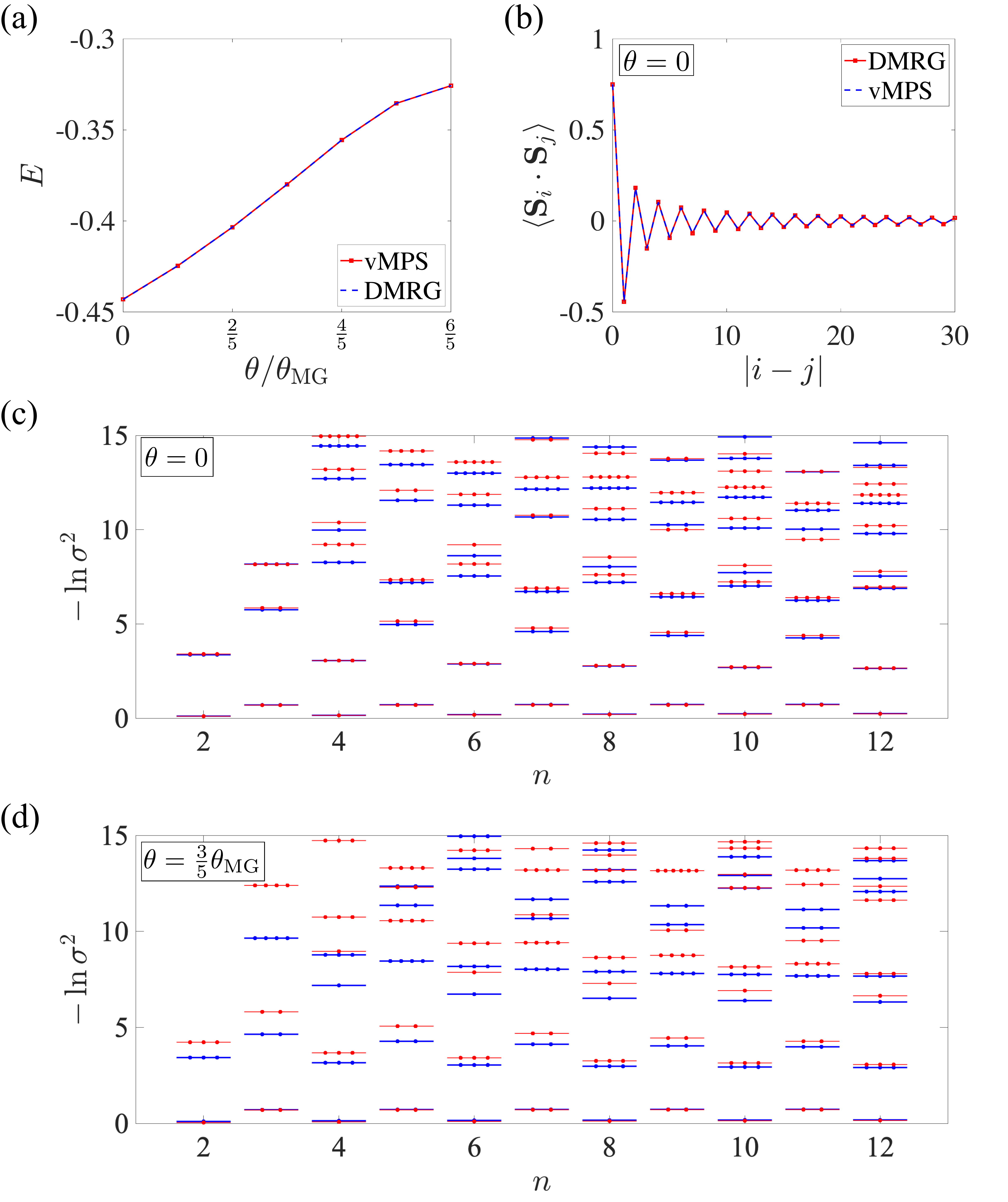} 
\caption{(a) Optimized energy of the $D=36$ vMPS and the DMRG
state as a function of the mixing angle $\theta$. (b) Spin-spin correlation function at $\theta =0$ (Heisenberg model) from vMPS and DMRG. (c)-(d) Bond-by-bond entanglement spectra obtained at the $(n, n+1)$ bond from the vMPS (black) and the DMRG (red) states for $\theta= 0$ and $\theta = (3/5)\theta_{\rm MG}$. The number of dots representing the degeneracy of each spectrum agrees well between vMPS and DMRG. ES is defined as minus the natural logarithm of the singular values of the density matrix.}
\label{fig:2}
\end{figure}

The $D=36$ vMPS wave function works well for capturing the ground states of the nearest and the next-nearest (NNN) neighbor spin-1/2 model
\ba H [\theta]  = \sum_i \Bigl[ \cos \theta ( {\v S}_i \cdot {\v S}_{i+1} )  + \sin \theta ( {\v S}_i \cdot {\v S}_{i+2} ) \Bigr] .  \label{eq:3.1}  \ea
The Heisenberg and the MG Hamiltonians are recovered at $\theta = 0$ and $\theta_{\rm MG}=\tan^{-1} (1/2)$, respectively. The $D=36$ vMPS optimization over a range of $\theta$ was performed with excellent energy comparison to the DMRG as shown in Fig. \ref{fig:2}(a), with the difference being no more than that at $\theta=0$. The spin-spin correlation function also agrees well between vMPS and DMRG as shown in Fig. \ref{fig:2}(b) at $\theta =0$, with similar agreement at other $\theta$'s (not shown). A closed chain of size $L = 2000$ was used for the vMPS optimization. Adopting the open boundary condition, one can also compute the entanglement spectrum (ES) at a given bond cut between sites $(n,n+1)$. Such bond-by-bond ES calculations are shown in Fig. \ref{fig:2}(c) for $\theta =0$ and \ref{fig:2}(d) for $\theta=(3/5)\theta_{\rm MG}$. The Heisenberg model is known to be integrable, and its ES levels reflect the conformal tower structure of the underlying critical theory~\cite{kim16}. We obtain identical degeneracy structures in the low-lying ES from both vMPS and DMRG for each bond cut regardless of whether the model is integrable $(\theta=0)$ or not ($\theta= (3/5)\theta_{\rm MG}$), further attesting to the validity of the vMPS wave function. In order to avoid breaking the SU(2) symmetry of the ansatz, we used the boundary vector $[b]_{j,j'}=\delta_{j0}\delta_{j'0}$ in both DMRG and vMPS calculations of the ES. 

A striking feature of the spin-1/2 chain is the existence of a spin-1/2 excitation known as the spinon. The physical picture of a spinon in the case of MG model is that of a free spin-1/2 flanked on either side by $|{\rm MG}_e \rangle$ and $|{\rm MG}_o \rangle$. In other words, the spinon is a ``domain wall" separating the two degenerate ground states. The excitation spectrum of the spinon for the MG model~\cite{shastry81} and the Heisenberg model~\cite{cloizeaux62} are known exactly. For other values of $\theta$ in the NNN model there are approximate formulas for the spinon spectrum~\cite{brehmer98,roux14}. The spinon spectrum was investigated numerically using the exact diagonalization method~\cite{affleck98,okunishi01}. Following the MG spinon construction, we can construct the one-spinon wave function by introducing the spinon site tensor $[S]_{ii',jj'}^s= v^s \delta_{i0} \delta_{j0}\delta_{i' 0}\delta_{j' 0}$, where $v^s$ is a coherent-state wave function representing a single spin, e.g. $(v^\uparrow, v^\downarrow) =(1 ,0)$ for spin-up and $(0, 1)$ for spin-down, respectively. Thus, the spinon vMPS is proposed to have the form 
\ba
|i\rangle_{\rm sp} = ( \prod_{j < i} \overline{T}^{s_j} ) \overline{S}^{s_{i}} ( \prod_{j> i} \overline{T}^{s_j } ) . \label{eq:spinon}\ea
As with the spinon in the MG model, a free spin-1/2 at $i$ ($[\overline{S}]_{ii',jj'}^{s} \equiv [S^s B]_{ii', jj}
=v^s\delta^0_{ij}\delta^0_{i'j'}$) becomes a domain wall between the two disconnected ground-state vMPS's. Projecting all virtual indices to the spin-0 sector guarantees that spins in the left-hand side and right-hand side of the $S$-tensor are completely disentangled. One can construct a slightly different $S$-tensor that allows finite entanglement between two sides. However, we have confirmed that the spinon wave function constructed in that way does not reproduce the expected spectrum well.

The momentum eigenstate $|k\rangle$ follows from the Fourier sum $|k\rangle = \sum_i e^{ik x_i}  |i\rangle_{\rm sp}$, and its energy from $\epsilon (k) = \langle k | H[\theta ] |k \rangle / \langle k | k \rangle$. The ground-state tensor $T^s$ had already been optimized for the Hamiltonian $H[\theta]$ and no additional effort was made to further optimize the spinon vMPS [Eq. (\ref{eq:spinon})], thus neglecting the screening of the isolated spin-1/2 by the neighboring spins. An odd number of sites ($L=2001$) is adopted in the calculation to accommodate a single spinon excitation in the vMPS. 

\begin{figure}[htb]
\includegraphics[width=0.48\textwidth]{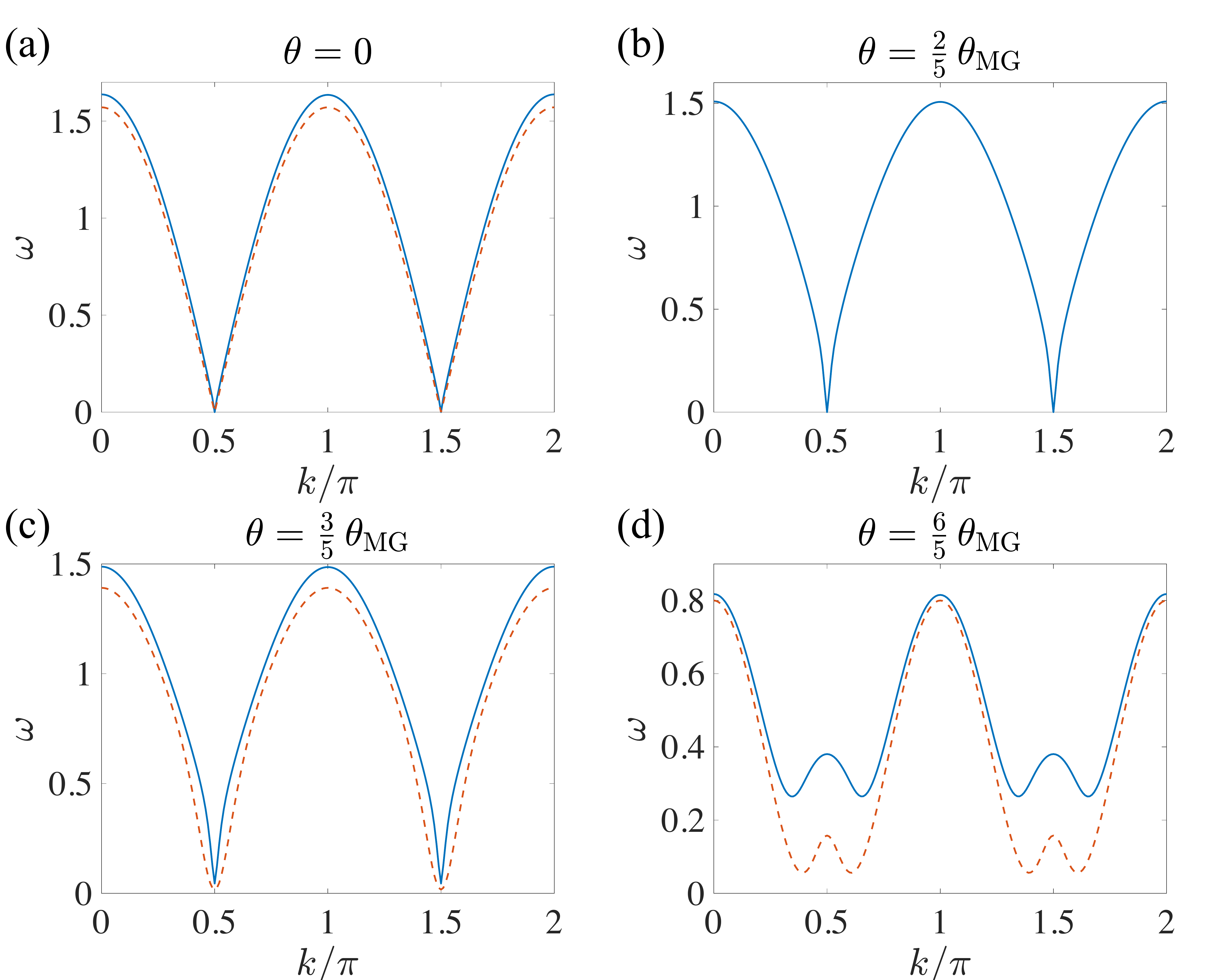} 
\caption{Spinon excitation spectrum at several values of $\theta$ in the NNN model. The $\theta=0$ spectrum (blue) matches the exact result $\epsilon (k) =  (\pi/2) |\cos k|$ (dotted). The fitting curves (dotted) in (c)-(d) are from Refs.~\cite{brehmer98,roux14}.   }
\label{fig:3}
\end{figure}

Figure \ref{fig:3} shows the spinon excitation spectrum at several $\theta$ values of the NNN Hamiltonian. At $\theta=0$, the calculated spectrum coincides very well with the exact one, $\epsilon^{\rm exact} (k) = (\pi/2) |\cos k|$ over the whole momentum range. The well-known spinon spectrum at the MG point~\cite{shastry81},  $\epsilon(k) = \cos \theta_{\rm MG} \times ( {5\over 8} +\frac{1}{2} \cos 2k) $ is recovered at $\theta=\theta_{\rm MG}$ (not shown). The spinon spectra at several other $\theta$ are shown in Fig. \ref{fig:3}. 
The NNN model is known to undergo a gap-closing transition at $\theta_c =  \tan^{-1}(0.241167) \approx 0.5104\times \theta_{\rm MG}$~\cite{haldane82,nomura92,eggert96}. Estimating the gap size from vMPS for $\theta> \theta_c$ or proving the gaplessness of the excitation for $\theta < \theta_c$, however, proved to be extremely challenging. Note that the spinon vMPS calculation is performed on a large lattice, $L=2001$, while the accuracy of the vMPS energy is in the first four digits. As a result, the {\it total energy} of the vMPS state becomes uncertain at around the same value as the expected gap size itself. The uncertainty in the DMRG energy adds to the conundrum. Performed on an open chain, the energy density of the DMRG wave function varies slightly from site to site. In obtaining the two curves in Fig. \ref{fig:3}(a)-(b) we simply subtracted the minimum value of the spinon energy $\epsilon (k)$. Subtracting the exactly DMRG values would create a small energy gap $\sim  0.2$ for $\theta =0$ and an invisibly smal one for $\theta = (2/5)\theta_{\rm MG}$. In Figs. \ref{fig:3}(c)-(d), actual DMRG ground state energy was subtracted. The double-dip structure in the spinon dispersion for $\theta > \theta_{\rm MG}$ was seen in earlier study~\cite{roux14}. 

A number of papers has addressed the ground state wave function of the spin-1/2 Heisenberg model in the MPS form~\cite{alcaraz03,alcaraz06,katsura10,korepin12}. 
Our vMPS is derived from a completely different perspective that reflects the RVB picture faithfully and gives quantitatively good fit to the exact wave function. The spinon excitation is understood simply as the creation of an isolated spin-1/2 between the two domains in the vMPS picture. Such a picture results in a quantitative replication of the spinon spectrum for the Heisenberg model. 

\acknowledgments J. H. H. was supported by Samsung Science and Technology Foundation under Project Number SSTF-BA1701-07. H.-Y.L. was supported by a Korea University Grant and National Research Foundation of Korea\,(NRF-2020R1I1A3074769).

\bibliography{SC}
\end{document}


\title{Supplementary Information for ``Matrix Product Wave Function of the Ground State and Elementary Excitation in the Spin-1/2 Chain''}

\author{Jintae Kim}
\affiliation{Department of Physics, Sungkyunkwan University, Suwon 16419, Korea}
\author{Minsoo Kim}
\affiliation{Department of Physics, Sungkyunkwan University, Suwon 16419, Korea}
\author{Pramod Padmanabhan}
\affiliation{Department of Physics, Sungkyunkwan University, Suwon 16419, Korea}
\author{Jung Hoon Han}
\email[Electronic address:$~~$]{hanjemme@gmail.com}
\affiliation{Department of Physics, Sungkyunkwan University, Suwon 16419, Korea}
\author{Hyun-Yong Lee}
\email[Electronic address:$~~$]{hyunyong@korea.ac.kr}
\affiliation{Department of Applied Physics, Graduate School, Korea University, Sejong 30019, Korea}
\affiliation{Division of Display and Semiconductor Physics, Korea University, Sejong 30019, Korea}
\affiliation{Interdisciplinary Program in E$\cdot$ICT-Culture-Sports Convergence, Korea University, Sejong 30019, Korea}

\date{\today}
\begin{abstract}
In this supplementary information we describe various technical details which were omitted in the main text. 
\end{abstract}  
\maketitle
\begin{widetext}

\section{Matrix Expressions of tensors}

The tensors we used for energy calculation is presented in terms of matrix. All of them are block diagonal matrices and $1\times 1$, $2 \times 2$, $3 \times 3$ diagonal blocks correspond to spin-$0$, $1/2$, $1$ sector, respectively.
\begin{align}
[{\rm CG}_{\frac{1}{2},\frac{1}{2}}^0]& =
\begin{pmatrix}
\begin{array}{c|cc|ccc}
0 & 0 & 0 & 0 & 0 & 0 \\ \hline
0 & 0 & \frac{1}{\sqrt{2}} & 0 & 0 & 0 \\
0 & -\frac{1}{\sqrt{2}} & 0 & 0 & 0 & 0 \\ \hline
0 & 0 & 0 & 0 & 0 & 0 \\
0 & 0 & 0 & 0 & 0 & 0 \\
0 & 0 & 0 & 0 & 0 & 0
\end{array}
\end{pmatrix}&
[{\rm CG}_{\frac{1}{2},1}^\frac{1}{2}]^{s=1}& =
\begin{pmatrix}
\begin{array}{c|cc|ccc}
0 & 0 & 0 & 0 & 0 & 0 \\ \hline
0 & 0 & 0 & 0 & -\frac{1}{\sqrt{3}} & 0 \\
0 & 0 & 0 & \sqrt{\frac{2}{3}} & 0 & 0 \\ \hline
0 & 0 & 0 & 0 & 0 & 0 \\
0 & 0 & 0 & 0 & 0 & 0 \\
0 & 0 & 0 & 0 & 0 & 0
\end{array}
\end{pmatrix}&
[{\rm CG}_{1,\frac{1}{2}}^\frac{1}{2}]^{s=1}& =
\begin{pmatrix}
\begin{array}{c|cc|ccc}
0 & 0 & 0 & 0 & 0 & 0 \\ \hline
0 & 0 & 0 & 0 & 0 & 0 \\
0 & 0 & 0 & 0 & 0 & 0 \\ \hline
0 & 0 & \sqrt{\frac{2}{3}} & 0 & 0 & 0 \\
0 & -\frac{1}{\sqrt{3}} & 0 & 0 & 0 & 0 \\
0 & 0 & 0 & 0 & 0 & 0
\end{array}
\end{pmatrix}\nn
\left[{\rm CG}_{1,1}^0\right]& =
\begin{pmatrix}
\begin{array}{c|cc|ccc}
0 & 0 & 0 & 0 & 0 & 0 \\ \hline
0 & 0 & 0 & 0 & 0 & 0 \\
0 & 0 & 0 & 0 & 0 & 0 \\ \hline
0 & 0 & 0 & 0 & 0 & \frac{1}{\sqrt{3}} \\
0 & 0 & 0 & 0 & -\frac{1}{\sqrt{3}} & 0 \\
0 & 0 & 0 & \frac{1}{\sqrt{3}} & 0 & 0
\end{array}
\end{pmatrix}&
[{\rm CG}_{\frac{1}{2},1}^\frac{1}{2}]^{s=2}& =
\begin{pmatrix}
\begin{array}{c|cc|ccc}
0 & 0 & 0 & 0 & 0 & 0 \\ \hline
0 & 0 & 0 & 0 & 0 & -\sqrt{\frac{2}{3}} \\
0 & 0 & 0 & 0 & \frac{1}{\sqrt{3}} & 0 \\ \hline
0 & 0 & 0 & 0 & 0 & 0 \\
0 & 0 & 0 & 0 & 0 & 0 \\
0 & 0 & 0 & 0 & 0 & 0
\end{array}
\end{pmatrix}&
[{\rm CG}_{1,\frac{1}{2}}^\frac{1}{2}]^{s=2}& =
\begin{pmatrix}
\begin{array}{c|cc|ccc}
0 & 0 & 0 & 0 & 0 & 0 \\ \hline
0 & 0 & 0 & 0 & 0 & 0 \\
0 & 0 & 0 & 0 & 0 & 0 \\ \hline
0 & 0 & 0 & 0 & 0 & 0 \\
0 & 0 & \frac{1}{\sqrt{3}} & 0 & 0 & 0 \\
0 & -\sqrt{\frac{2}{3}} & 0 & 0 & 0 & 0
\end{array}
\end{pmatrix}\nn
\end{align}

\section{Generalization of Tensors}
The $D=100$ tensor $T=T_1+T_1'+T_2+T_2'+T_3+T_3'+T_4+T_4'+T_5+T_5'+T_6+T_6'$ which is constructed by two $0 \oplus \frac{1}{2} \oplus 1 \oplus \frac{3}{2}$ virtual spaces and the $D=100$ bond tensor $B$ are following:
\ba
\left[T_1 \right]^s_{ii', jj'}  &=&  \delta_{i0} \delta_{js} [ \chi_1 ]_{i'j' } \quad~~~
\left[T_1' \right]^s_{ii', jj'}  =  \delta_{is} \delta_{j0} [\chi_{1'}]_{i' j'} \quad~~~~~
\left[ T_2 \right]^s_{ii', jj' } = [\chi_2]_{ij} \delta_{i' 0 } \delta_{j' s} \quad~~~
\left[ T_2' \right]^s_{ii', jj' } = [\chi_{2'}]_{i j}\delta_{i' s } \delta_{j' 0} \nn
\left[ T_3 \right]^s_{ii', jj' } &=&  [ {\rm CG}_{ {1\over 2} 1}^{ {1\over 2}}  ]^s_{i j} [\chi_3]_{i' j'}~~ \left[ T_3' \right]^s_{ii', jj' } =  [ {\rm CG}_{1 {1\over 2}}^{ {1\over 2}}  ]^s_{i j} [\chi_{3'}]_{i' j'} \quad~
\left[ T_4 \right]^s_{ii', jj' } = [\chi_4]_{i j} [ {\rm CG}_{{1\over 2} 1}^{{1\over2}} ]^s_{i'j' }~~ \left[ T_4' \right]^s_{ii', jj' } = [\chi_{4'}]_{i j} [ {\rm CG}_{1 {1\over 2} }^{{1\over2}} ]^s_{i'j' }\nn
\left[ T_5 \right]^s_{ii', jj' } &=&  [ {\rm CG}_{1 {3\over 2} }^{{1\over2}} ]^s_{ij }[\chi_5]_{i' j'}~~ \left[ T_5' \right]^s_{ii', jj' } =  [ {\rm CG}_{{3\over 2} 1}^{{1\over2}} ]^s_{ij} [\chi_{5'}]_{i' j'} \quad~
\left[ T_6 \right]^s_{ii', jj' } = [\chi_6]_{i j} [ {\rm CG}_{1 {3\over 2}}^{{1\over2}} ]^s_{i'j' }~~ \left[ T_6' \right]^s_{ii', jj' } = [\chi_{6'}]_{i j} [ {\rm CG}_{{3\over 2} 1}^{{1\over2}} ]^s_{i'j' }\nn
&&~~~~~~~~~~~~~~~~~~~~\left[\chi_{k}\right]_{ij} = \left( x_{k,0} \oplus x_{k,\frac{1}{2}} [ {\rm CG}_{ {1\over 2} {1\over 2}}^0] \oplus x_{k,1} [ {\rm CG}_{ {1} {1}}^0] \oplus x_{k,\frac{3}{2}} [ {\rm CG}_{ {\frac{3}{2}} \frac{3}{2}}^0] \right)_{ij}\nn
&&~~~~~~~~~~~B_{ii', jj' } =  ( 1 \oplus  [ {\rm CG}_{ {1\over 2} {1\over 2}}^0] \oplus  [ {\rm CG}_{1 1}^0 ] \oplus  [ {\rm CG}_{\frac{3}{2} \frac{3}{2}}^0 ])_{ij} \otimes ( 1 \oplus  [ {\rm CG}_{ {1\over 2} {1\over 2}}^0] \oplus  [ {\rm CG}_{1 1}^0 ] \oplus  [ {\rm CG}_{\frac{3}{2} \frac{3}{2}}^0 ])_{i'j'}.
\ea
\end{widetext}

\bibliography{SC}